\documentclass[12pt]{article}
\usepackage{epsf}
\usepackage[utf8]{inputenc}
\usepackage[english]{babel}

\begin{document}

\title{
{\bf Elastic diffractive scattering of nucleons at ultra-high energies}}
\author{A.A. Godizov\thanks{E-mail: anton.godizov@gmail.com}\\
{\small {\it Institute for High Energy Physics, 142281 Protvino, Russia}}}
\date{}
\maketitle

\begin{abstract}
A simple Regge-eikonal model with the eikonal represented as a single-reggeon-exchange term is applied to description of the nucleon-nucleon elastic 
diffractive scattering at ultra-high energies. The range of validity of the proposed approximation is discussed. The model predictions for the proton-proton 
cross-sections at the collision energy 14 TeV are given.
\end{abstract}

\section*{Introduction}

The fraction of the elastic diffractive scattering events in the total number of the $pp$ collision events at the collision energy 7-8 TeV is about 25\% \cite{totemdiff}. 
At higher energies it is expected to be even higher. Hence, understanding of the physical pattern of the small-angle elastic scattering of hadrons is indispensable for 
general understanding of the strong interaction at ultra-high energies. However, the special status of diffractive studies at high-energy colliders is determined by the fact 
that diffraction of hadrons takes place due to interaction at large distances. Indeed, the transverse size of the hadron interaction region can be estimated through straight 
application of the corresponding Heisenberg uncertainty relation to the experimental elastic angular distributions. For example, at the SPS, Tevatron, and LHC energies it is 
of order 1 fm. Therefore, exploitation of perturbative QCD for treatment of hadronic diffraction is disabled. 

The absence of exact theory leads to the emergence of numerous phenomenological models with very different underlying physics (the references to various models 
of the nucleon-nucleon elastic diffraction can be found in mini-review \cite{godizov}). The co-existence of a large number of rather complicated (and, often, incompatible) 
models points to the relevance of the question if construction of a much simpler (but adequate) approximation is possible. The word ``adequate'' in the last sentence 
implies as the theoretical correctness, so the satisfactory description of the high-energy evolution of the diffractive pattern.

The aim of this work is to demonstrate that a very simple and physically transparent description can be provided in the framework of the well-known Regge-eikonal approach 
\cite{collins} which originates from the synthesis of Regge theory and quasi-potential approximation. 

\section*{A model for high-energy elastic diffraction of nucleons}

The Regge-eikonal approach to description of the elastic scattering of hadrons exploits the eikonal representation of the nonflip scattering amplitude: 
\begin{equation}
\label{eikrepr}
T_{el}(s,t) = 4\pi s\int_0^{\infty}db^2J_0(b\sqrt{-t})\frac{e^{2i\delta(s,b)}-1}{2i}\,,
\end{equation}
$$
\delta(s,b) = \frac{1}{16\pi s}\int_0^{\infty}d(-t)J_0(b\sqrt{-t})\delta(s,t)\,,
$$
where $s$ and $t$ are the Mandelstam variables, $b$ is the impact parameter, and eikonal $\delta(s,t)$ is the sum of single-reggeon-exchange terms:
\begin{equation}
\label{ppeik}
\delta(s,t) = \sum_n\xi(\alpha_n^+(t))\Gamma_n^{(1)+}(t)\Gamma_n^{(2)+}(t)\left(\frac{s}{s_0}\right)^{\alpha_n^+(t)}
\mp\,\sum_n\xi(\alpha_n^-(t))\Gamma_n^{(1)-}(t)\Gamma_n^{(2)-}(t)\left(\frac{s}{s_0}\right)^{\alpha_n^-(t)},
\end{equation}
where $\alpha_n^+(t)$ and $\alpha_n^-(t)$ are the $C$-even and $C$-odd Regge trajectories, $\Gamma_n^{(i)}(t)$ are the corresponding reggeon form-factors of the colliding 
particles, $s_0 = 1$ GeV$^2$, $\xi(\alpha(t))$ are the so-called reggeon signature factors, 
$\xi(\alpha(t))=i+{\rm tg}\frac{\pi(\alpha(t)-1)}{2}$ for even reggeons,\footnote{Namely, even (odd) Regge trajectories are the analytic continuations of the corresponding 
even-spin (odd-spin) resonance spectra.} $\xi(\alpha(t))=i-{\rm ctg}\frac{\pi(\alpha(t)-1)}{2}$ for odd reggeons, and the sign ``$-$'' (``$+$'') before the $C$-odd reggeon 
contributions corresponds to the particle-particle (particle-antiparticle) interaction. More detailed discussion of the eikonal Regge approximation can be found in 
\cite{collins} or, for instance, \cite{kisselev}. An important advantage of this approach is that it allows to satisfy the Froissart-Martin bound \cite{froissart} 
explicitly. 

Let us consider the ultimate case of the nucleon-nucleon diffraction at ultra-high energies, where the eikonal can be approximated by the only Regge pole term corresponding 
to the leading even and $C$-even reggeon (called ``pomeron'' or ``soft pomeron'' in literature):
\begin{equation}
\label{eikphen}
\delta(s,t) = \delta_{\rm P}(s,t) \equiv \left(i+{\rm tg}\frac{\pi(\alpha_{\rm P}(t)-1)}{2}\right){\Gamma_{\rm P}}^2(t)\left(\frac{s}{s_0}\right)^{\alpha_{\rm P}(t)},
\end{equation}
where $\alpha_{\rm P}(t)$ is the Regge trajectory of pomeron and $\Gamma_{\rm P}(t)$ is the pomeron form-factor of nucleon.

In the current stage of development, QCD provides no useful information about the behavior of $\alpha_{\rm P}(t)$ and $\Gamma_{\rm P}(t)$ in the diffraction domain 
($0<-t<2$ GeV$^2$), though it was argued \cite{kearney} that $\alpha_{\rm P}(t)>1$ at $t<0$ and 
\begin{equation}
\label{gluon}
\lim_{t\to -\infty}\alpha_{\rm P}(t) = 1\,.
\end{equation}
Therefore, we will use the simplest test parametrizations\footnote{One should not consider the analytic properties of parametrizations (\ref{pomeron}) seriously. True 
Regge trajectories and reggeon form-factors have much more complicated analytic structure. Nonetheless, at {\it negative} values of the argument, they can be approximated 
by simple monotonic test functions.}
\begin{equation}
\label{pomeron}
\alpha_{\rm P}(t) = 1+\frac{\alpha_{\rm P}(0)-1}{1-\frac{t}{\tau_a}}\,,\;\;\;\Gamma_{\rm P}(t) = \frac{\Gamma_{\rm P}(0)}{\left(1-\frac{t}{\tau_g}\right)^2}\,.
\end{equation}

To obtain the angular distribution, one should substitute (\ref{pomeron}) into (\ref{eikphen}), then, using representation (\ref{eikrepr}), calculate the scattering 
amplitude, and, at last, substitute it into the expression for differential cross-section: 
\begin{equation}
\label{diffsech}
\frac{d\sigma_{el}}{dt} = \frac{|T_{el}(s,t)|^2}{16\pi s^2}\,.
\end{equation}

To fit the model parameters, we restrict ourselves by the SPS, Tevatron, and LHC energies ($\sqrt{s}>$ 500 GeV) and small transfers of momentum 
(0.005 GeV$^2<-t<$ 2 GeV$^2$), since the test parametrization (\ref{pomeron}) of form-factor $\Gamma_{\rm P}(t)$ is too stiff and does not allow to provide a satisfactory 
description of the data in both the diffraction domain and the hard scattering region simultaneously, while at the ISR energies the influence of secondary reggeons becomes 
significant enough to distort the diffractive pattern crucially (the restriction in the collision energy from below is discussed in detail in the following section).
\begin{figure}[ht]
\vskip -0.5cm
\epsfxsize=8.2cm\epsfysize=8.2cm\epsffile{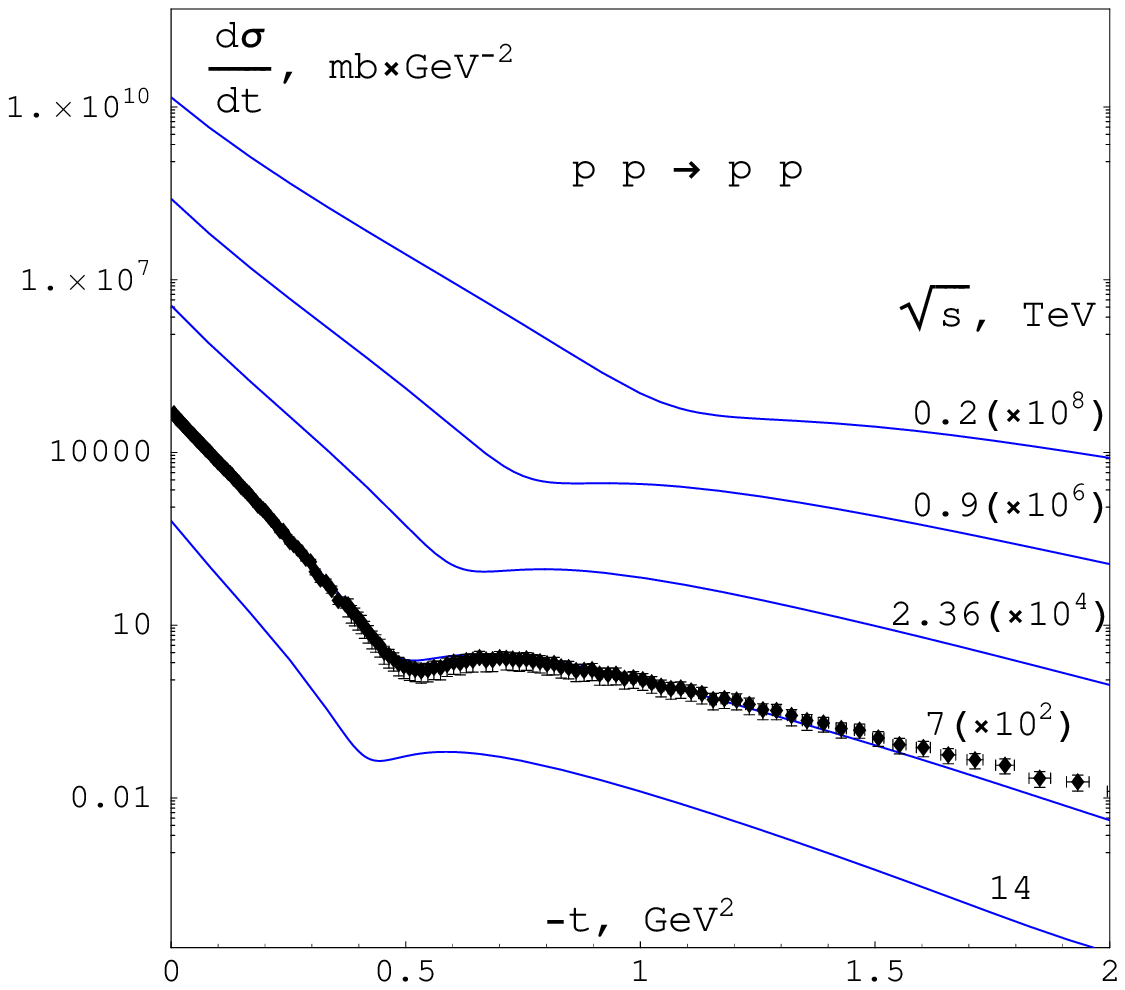}
\vskip -8.25cm
\hskip 8.5cm
\epsfxsize=8.2cm\epsfysize=8.2cm\epsffile{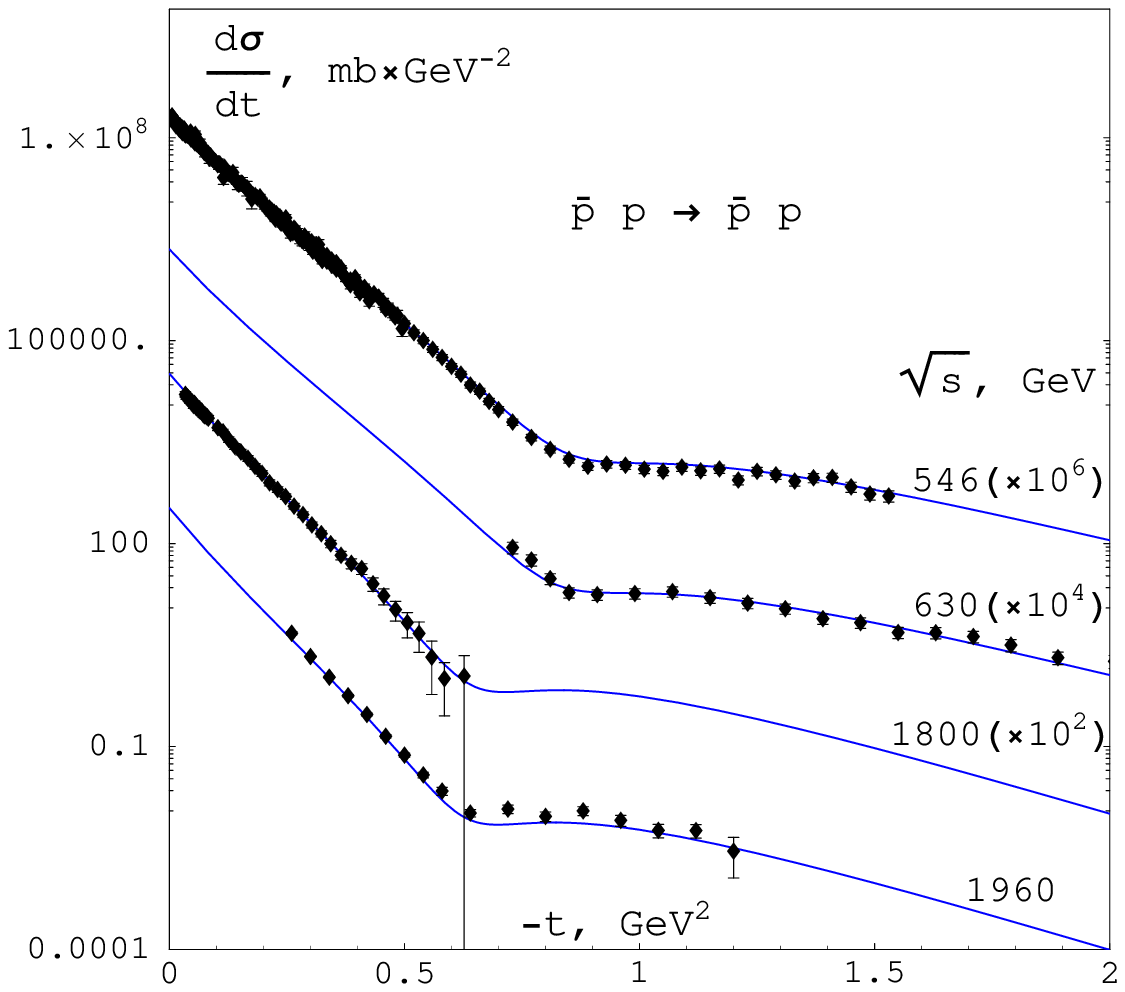}
\end{figure}
\begin{figure}[ht]
\vskip -0.85cm
\hskip 0.65cm
\epsfxsize=7.5cm\epsfysize=7.5cm\epsffile{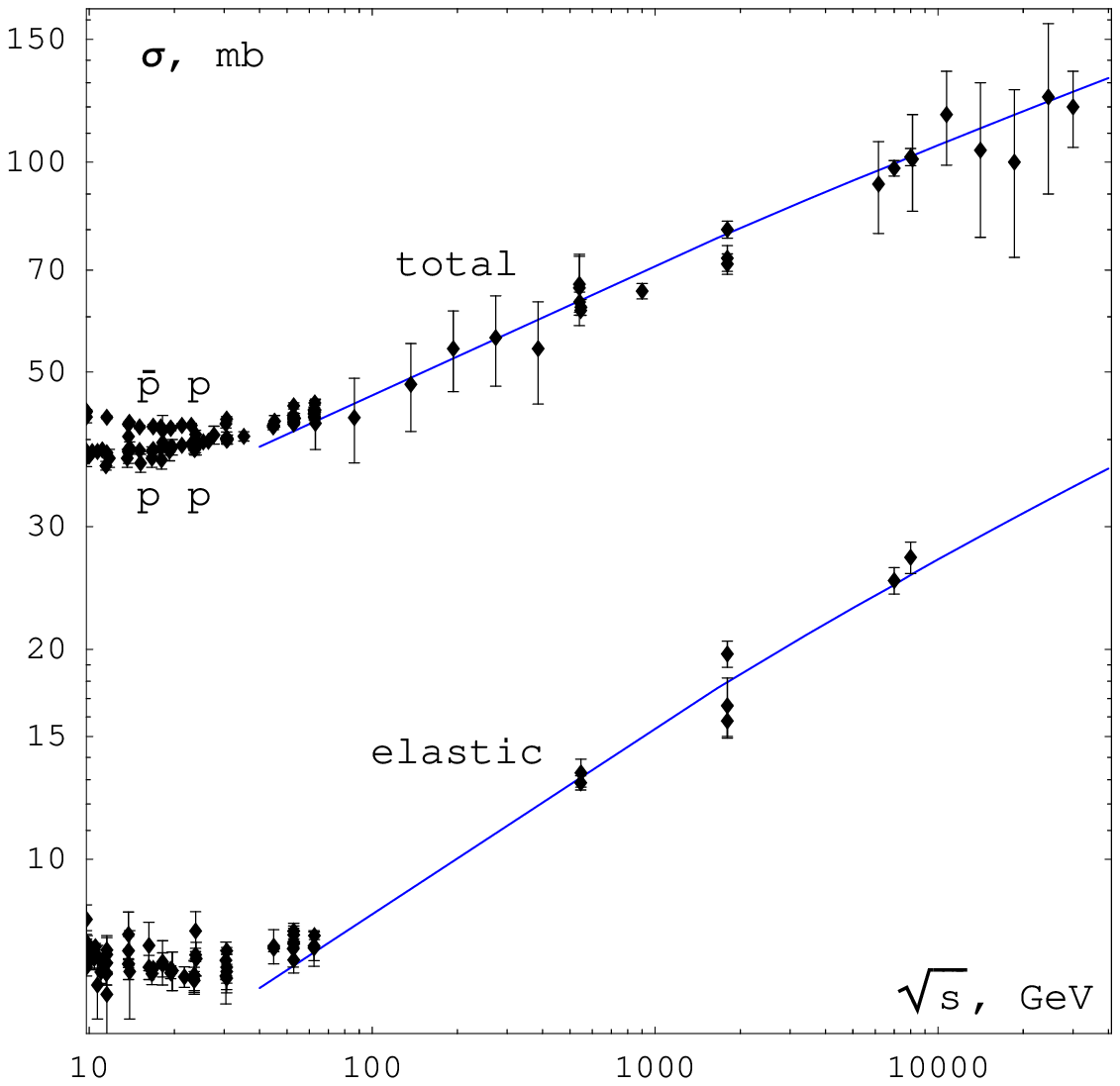}
\vskip -7.55cm
\hskip 9.1cm
\epsfxsize=7.5cm\epsfysize=7.5cm\epsffile{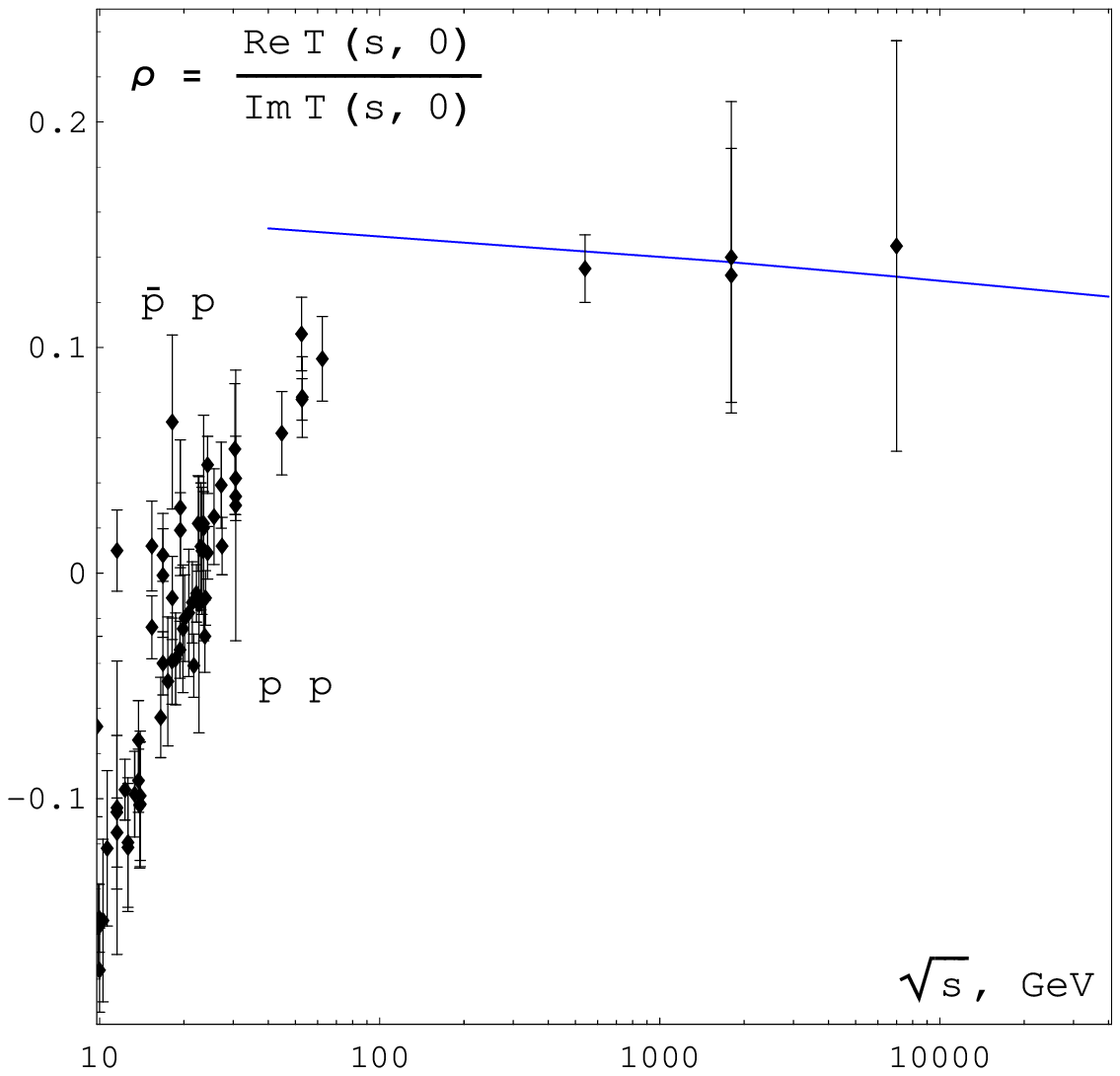}
\caption{Description of the nucleon-nucleon elastic diffraction observables at ultra-high values of the collision energy.}
\label{pp}
\end{figure}

For verification of the model, we used the set of data on angular distributions collected by J.R. Cudell, A. Lengyel, and E. Martynov at \cite{cudell}. The original data 
can be found in \cite{totemdiff,diffexp,total}. The results of the model application to description of the nucleon-nucleon elastic diffraction are presented in 
Fig. \ref{pp} and Tabs. \ref{tab1}, \ref{tab2}, and \ref{tab3}. As well, in Fig. \ref{pp} and Tab. \ref{tab3} there can be found the predictions for the $pp$ differential, 
elastic and total cross-sections at various values of the collision energy, including $\sqrt{s}=14$ TeV. 

The model description of the nucleon-nucleon diffractive pattern evolution from 0.5 TeV to 7 TeV of the collision energy is satisfactory\footnote{At $\sqrt{s}=546$ GeV, 
3 outlying points were excluded from the fitting procedure: at $t_1=-0.01375$ GeV$^2$, $t_2=-0.034$ GeV$^2$, and $t_3=-1.21$ GeV$^2$.}: $\chi^2/DoF\approx 0.97$. 

\begin{table}[ht]
\begin{center}
\begin{tabular}{|l|l|}
\hline
\bf Parameter          & \bf Value                  \\
\hline
$\alpha_{\rm P}(0)-1$  & $0.111\pm 0.017$             \\
$\tau_a$                  & $(0.47\pm 0.12)$ GeV$^2$     \\
$\Gamma_{\rm P}(0)$    & $ 7.43\pm 0.94$              \\
$\tau_g$                  & $(0.98\pm 0.12)$ GeV$^2$     \\
\hline
\end{tabular}
\end{center}
\caption{The parameter values for expressions (\ref{pomeron}), obtained via fitting to the differential cross-section data.}
\label{tab1}
\end{table}

\begin{table}[ht]
\begin{center}
\begin{tabular}{|l|l|l|}
\hline
$\sqrt{s}$, GeV                      & \bf Number of points &  $\chi^2$   \\
\hline
          546 ($\bar p\,p$)          & 228           &  222        \\
          630 ($\bar p\,p$)          & 17            &   13        \\
         1800 ($\bar p\,p$)          & 51            &   17        \\
         1960 ($\bar p\,p$)          & 17            &   52        \\
         7000 ($p\,p$)               & 161           &  151        \\
\hline
\bf Total & 474 & 455     \\
\hline
\end{tabular}
\end{center}
\caption{The quality of description of the data on the nucleon-nucleon angular distributions.}
\label{tab2}
\end{table}

\begin{table}[ht]
\begin{center}
\begin{tabular}{|l|l|l|}
\hline
$\sqrt{s}$, GeV   & $\sigma_{tot}$, mb &   $\sigma_{el}$, mb  \\
\hline
    62.5      & $ 42.6\pm 4.0$           &   $ 7.4\pm 1.1$        \\
   200        & $ 53.0\pm 3.5$           &   $10.2\pm 1.0$        \\
   546        & $ 63.8\pm 3.3$           &   $13.3\pm 0.9$        \\
  1800        & $ 79.0\pm 4.2$           &   $18.0\pm 1.1$        \\
  7000        & $ 99.6\pm 7.3$           &   $24.8\pm 2.2$        \\
  8000        & $101.8\pm 7.7$           &   $25.6\pm 2.3$        \\
 13000        & $110.3\pm 9.4$           &   $28.5\pm 2.9$        \\
 14000        & $111.7\pm 9.6$           &   $29.0\pm 3.0$        \\
\hline
\end{tabular}
\end{center}
\caption{Predictions for the $pp$ total and elastic cross-sections.}
\label{tab3}
\end{table}

\section*{The impact of secondary reggeons at the ISR energies}

\begin{figure}[ht]
\vskip -0.5cm
\begin{center}
\epsfxsize=11.5cm\epsfysize=11.5cm\epsffile{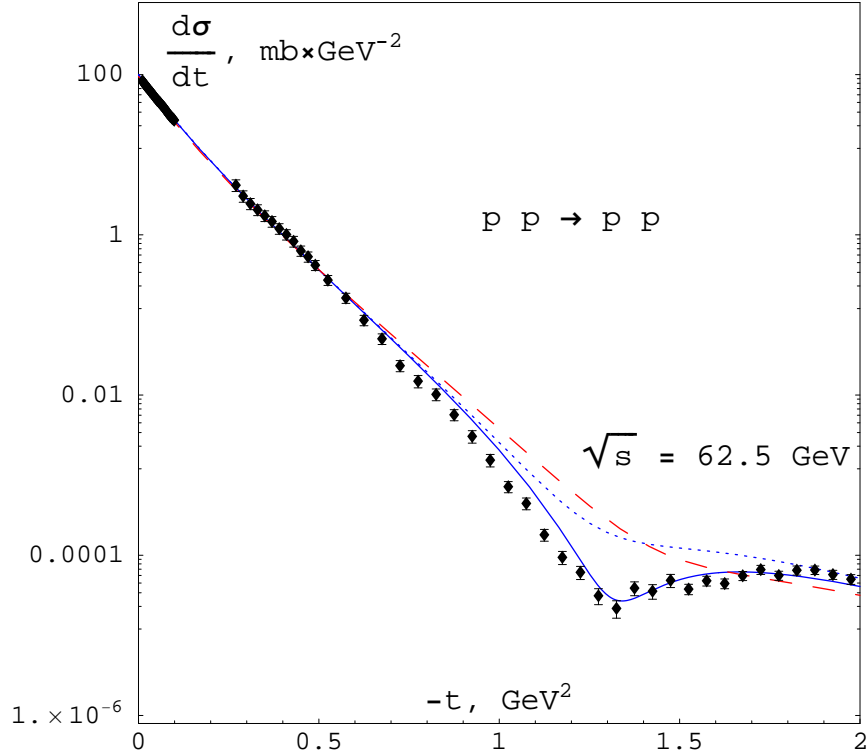}
\vskip -0.5cm
\caption{The angular distribution of the $pp$ elastic scattering at $\sqrt{s}=$ 62.5 GeV.}
\label{isr62}
\end{center}
\end{figure}

Let us turn to investigation of the proposed approximation validity (or invalidity) at lower energies. The model straight application to description of the $pp$ elastic 
scattering at $\sqrt{s}=$ 62.5 GeV \cite{isr62} reveals a huge discrepancy between the theoretical curve and the data (Fig. \ref{isr62}, the dashed line), particularly 
in the dip region. The fact of such a disagreement is not surprising, since the impact of secondary reggeon exchanges is expected to be significant at the ISR energies. 
Adding some test exponential term to the eikonal, $\delta(s,t)=\delta_{\rm P}(s,t)\to\delta_{\rm P}(s,t) + (i-1)\,\beta \, e^{\,b\,t}$ (here $s=$ (62.5 GeV)$^2$, 
$\beta = 6\cdot 62.5^2$, and $b = 4$ GeV$^{-2}$), one makes this disagreement much less catastrophic (Fig. \ref{isr62}, the solid line). The dotted line in Fig. \ref{isr62} 
corresponds to the replacement $\delta_{\rm P}(s,t)\to\delta_{\rm P}(s,t) + i\,\beta \, e^{\,b\,t}$ (in the regime Re$\,\delta(s,t)\ll$ Im$\,\delta(s,t)$, the dip 
position is determined by Im$\,\delta$, while the dip depth is determined by Re$\,\delta$).

The question is whether a comparable modification ({\it i.e.} a few percents increase of Im$\,\delta(s,t)$ and a 2-3 times decrease of Re$\,\delta(s,t)$) can be caused by 
the combined contribution of secondary reggeons into the eikonal at the $pp$ collision energy 62.5 GeV. Four secondary reggeons are expected to give a noticeable 
contribution at the ISR energies: two even and $C$-even reggeons, $f$ and $a$, and two odd and $C$-odd reggeons, $\omega$ and $\rho$. With account of these secondaries, 
the $pp$ scattering eikonal (\ref{ppeik}) takes the following form:
$$
{\rm Im}\,\delta(s,t) = {\rm Im}\,\delta_{\rm P}(s,t) + {\rm Im}\,\delta_{\rm R}(s,t) \equiv {\Gamma_{\rm P}}^2(t)\left(\frac{s}{s_0}\right)^{\alpha_{\rm P}(t)} + 
$$
$$
+ \left[{\Gamma_f}^2(t)\left(\frac{s}{s_0}\right)^{\alpha_f(t)} - {\Gamma_\omega}^2(t)\left(\frac{s}{s_0}\right)^{\alpha_\omega(t)} 
+ {\Gamma_a}^2(t)\left(\frac{s}{s_0}\right)^{\alpha_a(t)} - {\Gamma_\rho}^2(t)\left(\frac{s}{s_0}\right)^{\alpha_\rho(t)}\right]\,,
$$
\begin{equation}
\label{eikseco}
{\rm Re}\,\delta(s,t) = {\rm Re}\,\delta_{\rm P}(s,t) + {\rm Re}\,\delta_{\rm R}(s,t) \equiv
{\rm tg}\frac{\pi(\alpha_{\rm P}(t)-1)}{2}{\Gamma_{\rm P}}^2(t)\left(\frac{s}{s_0}\right)^{\alpha_{\rm P}(t)} + 
\end{equation}
$$
+ \left[{\rm tg}\frac{\pi(\alpha_f(t)-1)}{2}\,{\Gamma_f}^2(t)\left(\frac{s}{s_0}\right)^{\alpha_f(t)} 
+ {\rm ctg}\frac{\pi(\alpha_\omega(t)-1)}{2}\,{\Gamma_\omega}^2(t)\left(\frac{s}{s_0}\right)^{\alpha_\omega(t)} + \right.
$$
$$
\left. + \; {\rm tg}\frac{\pi(\alpha_a(t)-1)}{2}\,{\Gamma_a}^2(t)\left(\frac{s}{s_0}\right)^{\alpha_a(t)} 
+ {\rm ctg}\frac{\pi(\alpha_\rho(t)-1)}{2}\,{\Gamma_\rho}^2(t)\left(\frac{s}{s_0}\right)^{\alpha_\rho(t)}\right]\,.
$$

\begin{figure}[ht]
\epsfxsize=7.8cm\epsfysize=7.8cm\epsffile{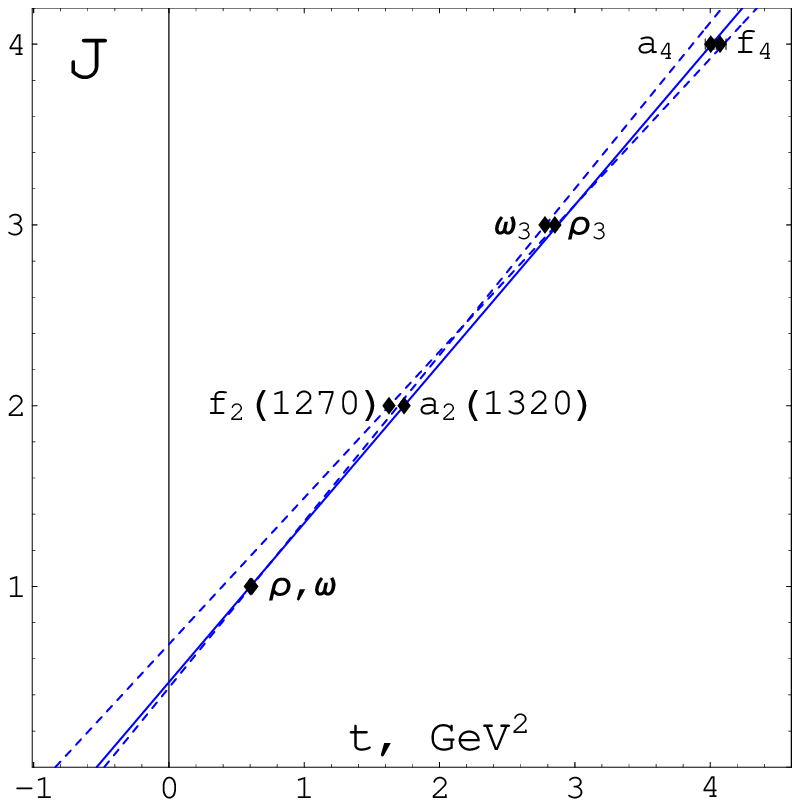}
\vskip -8cm
\hskip 8.5cm
\epsfxsize=8.2cm\epsfysize=8.2cm\epsffile{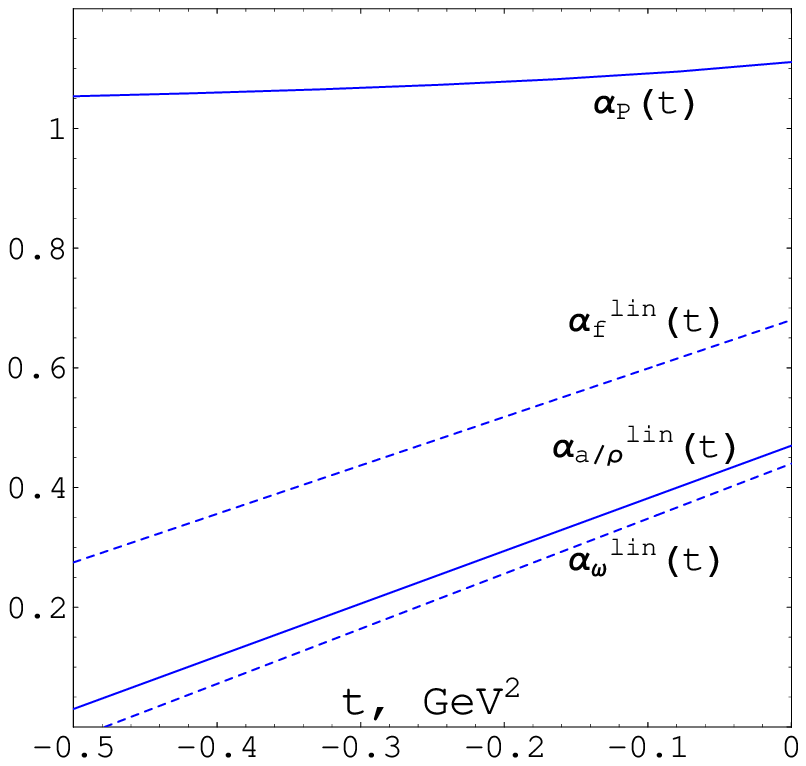}
\caption{The Chew-Frautschi plots of secondary reggeons in the resonance region (on the left) and in the diffractive scattering region (on the right):
$\alpha_f(t)=0.68+0.81t$, $\alpha_\omega(t)=0.44+0.92t$, $\alpha_{a/\rho}(t)=0.47+0.88t$.}
\label{reson}
\end{figure}

The secondary reggeon intercepts can be estimated with the help of their linear Chew-Frautschi plots\footnote{The resonance masses are determined from the equations like 
$\alpha(M^2-{\rm i} M\Gamma)=J$ (here $\alpha(t)$ is some Regge trajectory and $J$ is the resonance spin) which do not imply that ${\rm Re}\,\alpha(M^2)=J$ and 
${\rm Im}\,\alpha(M^2)=0$. The decay widths $\Gamma$ are only few times less than the corresponding masses $M$: for example, for $f_2(1270)$-meson 
$\frac{\Gamma_f}{M_f}\approx 0.15$, and for $\rho(770)$-meson $\frac{\Gamma_\rho}{M_\rho}\approx 0.19$. Consequently, the Chew-Frautschi plots should be considered just as 
very rough approximations to the true Regge trajectories in the resonance region.} (Fig. \ref{reson}): $\alpha_f(0)= 0.68$, $\alpha_\omega(0)= 0.44$, 
$\alpha_a(0)=\alpha_\rho(0)= 0.47$, and, thus, ${\rm tg}\frac{\pi(\alpha_f(0)-1)}{2}\approx -0.55$, ${\rm ctg}\frac{\pi(\alpha_\omega(0)-1)}{2}\approx -0.8$, 
${\rm tg}\frac{\pi(\alpha_a(0)-1)}{2}\approx -1.1$, ${\rm ctg}\frac{\pi(\alpha_\rho(0)-1)}{2}\approx -0.9$. Regarding the $t$-evolution of the leading $\bar qq$-reggeons, 
one should keep in mind that $\alpha_{\bar qq}(t)\sim\ln^{-1/2}(-t)$ at $t\to -\infty$ \cite{kwiecinski}. Hence, in the region $t<0$, the true secondary Regge trajectories 
should be essentially nonlinear and very different from the corresponding Chew-Frautschi plots. Therefore, for all of the mentioned secondaries, we expect the real parts 
of their signature factors to be negative at $t<0$ and to be, at least, several times higher in magnitude than the factor 
${\rm tg}\frac{\pi(\alpha_{\rm P}(t)-1)}{2}<{\rm tg}\frac{\pi(\alpha_{\rm P}(0)-1)}{2}\approx 0.2$. As well, it should be noted that in the $pp$ elastic scattering the 
contributions of $\omega$ and $\rho$ into ${\rm Im}\,\delta(s,t)$ partially compensate the contributions of $f$ and $a$, while in ${\rm Re}\,\delta(s,t)$ the 
corresponding terms have the same (negative) sign (see (\ref{eikseco})). Under such conditions, it seems quite natural that, at the $pp$ collision energy 62.5 GeV and low 
values of the transferred momentum, the combined contribution of secondary reggeons makes ${\rm Im}\,\delta(s,t)$ a few percents higher than ${\rm Im}\,\delta_{\rm P}(s,t)$ 
and makes ${\rm Re}\,\delta(s,t)$ two-three times lower than ${\rm Re}\,\delta_{\rm P}(s,t)$.

\begin{figure}[ht]
\vskip -0.5cm
\begin{center}
\epsfxsize=10cm\epsfysize=10cm\epsffile{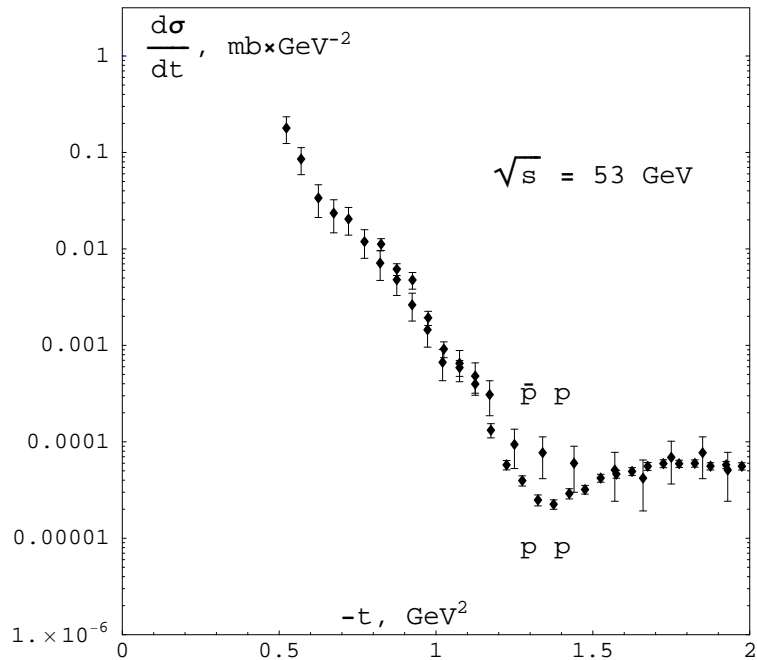}
\vskip -0.5cm
\caption{The experimental angular distributions of the $\bar pp$ \cite{isr53app} and $pp$ \cite{isr53pp} elastic scattering at $\sqrt{s}=$ 53 GeV.}
\label{isr53}
\end{center}
\end{figure}

In the framework of the discussed physical pattern, it is possible to explain the noticeable discrepancy between the $\bar pp$ and $pp$ differential cross-sections in some 
vicinity of the dip (Fig. \ref{isr53}) without any appeal to the notion of odderon (the leading $C$-odd reggeon with the intercept higher than 1.0). Due to the suppression 
of the pomeron term in ${\rm Re}\,\delta(s,t)$ by the factor ${\rm tg}\frac{\pi(\alpha_{\rm P}(t)-1)}{2}$, the relative contribution of $\omega$ and $\rho$ into 
${\rm Re}\,\delta(s,t)$ is much more significant than into ${\rm Im}\,\delta(s,t)$. These real terms are positive for $\bar pp$ scattering and negative for $pp$ scattering. 
Therefore, ${\rm Re}\,\delta(s,t)$ for $\bar pp$ scattering could be noticeably higher than for $pp$ scattering. As the angular distribution behavior in some small vicinity 
of the dip depends on ${\rm Re}\,\delta$ strongly, so, in this vicinity, the $\bar pp$ differential cross-section is expected to be appreciably larger than the $pp$ 
differential cross-section (and this is the very pattern which takes place in experiment). If the odderon exchange effect is really negligible, the splitting between the 
$\bar pp$ and $pp$ angular distributions should vanish at higher energies.

In view of the above-said, we infer that although the one-reggeon (pomeron) eikonal approximation to the nucleon-nucleon elastic scattering amplitude is invalid at 
the ISR energies, the deviations from the experimental data on the corresponding differential cross-sections are, in principle, consistent with the expected phenomenology 
of secondary reggeons at low negative values of $t$.

The influence of secondaries decreases fast with the collision energy growth. For instance, at $\sqrt{s}=$ 200 GeV the combined relative contribution of secondary reggeon 
exchanges is expected to be about 1-2 percents for ${\rm Im}\,\delta(s,t)$ and about 10-20 percents for ${\rm Re}\,\delta(s,t)$. Therefore, the model predictions of the $pp$ 
scattering observables at the RHIC energy (see Fig. \ref{pp}) could be adequate not only for $\sigma_{tot}$ and $\sigma_{el}$, but also for $\frac{d\sigma_{el}}{dt}$, 
excluding the very vicinity of the dip. In other words, at the collision energy about 200 GeV the nucleon-nucleon elastic diffraction passes into the pure pomeron-exchange 
regime.

\section*{Conclusions}

The practical benefit of the one-reggeon eikonal approximation is that, in the kinematic range of diffractive scattering, it allows to reduce the unknown complex function of 
two variables, $T(s,t)$, to 2 monotonic functions of one variable, $\alpha_{\rm P}(t)$ and $\Gamma_{\rm P}(t)$. In spite of its roughness\footnote{Certainly, for better 
description of experimental data, one could use more complicated and flexible parametrizations of $\alpha_{\rm P}(t)$ and, especially, of $\Gamma_{\rm P}(t)$ than stiff 
test expressions (\ref{pomeron}).}, the proposed phenomenological scheme provides a satisfactory description of the nucleon-nucleon elastic scattering at the collision 
energies 0.54 TeV $\le\sqrt{s}\le$ 7 TeV and transferred momenta squared 0.005 GeV$^2<-t<$ 2 GeV$^2$ and allows to give well-grounded predictions for the diffractive pattern 
at higher (the LHC) and lower (the RHIC) energies. 

Nonetheless, confirmation or discrimination of the model is possible after the forthcoming TOTEM measurements of the $pp$ differential cross-section at $\sqrt{s}=$ 14 
TeV (the analogous measurements at the RHIC are, as well, very desirable). In the case of confirmation, the one-reggeon eikonal approximation could become an effective tool 
for treatment of the nucleon-nucleon elastic diffraction at ultra-high energies, due to its salient simplicity and physical clearness.

Besides, the pomeron Regge trajectory and the pomeron form-factor of nucleon are used in the framework of the Regge approach to more complicated (than $2\to 2$) diffractive 
reactions: single diffraction ($p+p\to p+X$ or $\bar p+p\to \bar p+X$), central exclusive diffractive production of the Higgs boson ($p+p\to p+H+p$), {\it etc...} Therefore, 
the possibility of implicit extraction of $\alpha_{\rm P}(t)$ and $\Gamma_{\rm P}(t)$ from the experimental angular distributions of nucleon-nucleon elastic scattering is 
very important for raising the predictive efficiency of the reggeon models describing the energy evolution of the corresponding cross-sections.

\end{document}